\documentclass[aps,prd,preprint,superscriptaddress,showpacs,tightenlines]{revtex4}
\usepackage{graphics}
\usepackage{epsfig}
\usepackage{latexsym}
\usepackage{colordvi}

\begin{document}

\preprint{\parbox[b]{1in}{ \hbox{\tt PNUTP-04/A07} }}


\title{Light Scalar Decay in Diquark Chiral Effective Theory}

\author{Deog Ki Hong}
\email{dkhong@pusan.ac.kr} 
\author{Chaejun Song}
\email{chaejun@charm.phys.pusan.ac.kr}

\affiliation{Department of Physics, Pusan National University,
Busan 609-735, Korea}

\vspace{0.1in}

\date{\today}

\begin{abstract}
We calculate the decay rate of light scalar mesons, using a
diquark chiral effective theory,  recently proposed to describe
exotic hadrons. In the effective theory the light scalar mesons are
postulated to be bound states of diquark and anti-diquark.
We find our results
are in good agreement with experimental data. The axial couplings
of diquarks with pions and kaons are found to be quite small and
the perturbation is reliable. It shows that the diquark picture
captures the correct physics of light scalar mesons.
\end{abstract}

\pacs{12.39.-x, 13.25.Jx, 12.38.-t}

\maketitle
\section{Introduction}
Recent discovery of exotic baryons~\cite{Nakano:2003qx} revived the interest
in hadron spectroscopy, especially that of exotic hadrons.
Among several models for exotic baryons, promising one is the diquark
model~\cite{Jaffe:2003sg,Shuryak:2003zi}, which assumes a strong
correlation of two quarks in the color antitriplet
channel. In the Jaffe-Wilczek (JW) diquark model~\cite{Jaffe:2003sg},
the exotic baryons are
bound states of two scalar diquarks and one antiquark.
The salient features of JW model are the occurrence of low-dimensional multiplets
and small mass difference among the members in the multiplet, which were soon
supported by the NA49 experiment~\cite{Alt:2003vb}.
Furthermore, the extremely narrow decay
width of exotic baryons was naturally explained in the diquark
picture of exotic baryons, where the decay process
is suppressed by tunnelling among two diquarks~\cite{Hong:2004xn}.

As the existence of hadrons containing quarks more than three is
logically possible, its absence in the low-lying hadronic
spectroscopy has been a puzzle since the advent of the quark
model~\cite{Gell-Mann:nj}. In late seventies, Jaffe suggested that
the positive-parity scalar mesons of mass less than $1~{\rm GeV}$
might be bound states of multi quarks, forming a $SU(3)$
nonet~\cite{Jaffe:1976ig}. In his scheme, the scalar nonet is
postulated to be the non-exotic components of 4-quark states
($Q^2{\bar Q}^2$). However, the quark content of light scalar
mesons has been controversial and still is~\cite{VanBeveren:ea}.
Recently, it was argued that $\sigma$, the lightest member in the
scalar nonet, is not a pure $Q\bar Q$ state but mostly a
four-quark state, using the unitarity argument in the large $N_C$
analysis~\cite{Sannino:1995ik}. (See also a recent paper for the
diquark picture of scalar mesons~\cite{Maiani:2004uc}.)

In this paper, we assume that the scalar nonet is a bound sate of
a diquark and an anti-diquark and calculate the decay width of scalar nonet
to compare with the experimental data~\cite{Achasov:2000ku,Hagiwara:2002fs},
using the diquark chiral effective theory~\cite{Hong:2004xn}.

\section{Diquark chiral effective theory}
\label{dcet}

The success of quark model in hadron spectroscopy is understood in the
framework of the chiral quark effective theory ($\chi$QET), derived by Georgi and
Manohar~\cite{Manohar:1983md}. The $\chi$QET is an effective theory
of quantum chromodynamics (QCD) below the chiral symmetry breaking scale
$\Lambda_{{\chi}{\rm SB}}$ but above the confinement scale, $\Lambda_{\rm QCD}$.
The relevant degrees of freedom of the effective theory therefore contains quarks,
gluons and the Nambu-Goldstone (NG) bosons of spontaneously
broken $SU(3)_L\times SU(3)_R$ chiral symmetry.
The chiral quark effective Lagrangian is given as
\begin{equation}
{\cal L}_{{\chi}Q}=\bar\psi\left(i\!\not\!\! D\, +\not\! V\right)\psi +g_A\bar\psi
\not\!\! A\,\gamma_5\psi-m\bar\psi\psi
+{1\over2}f^2\,{\rm tr}\,\partial^{\mu}\Sigma^{\dagger}\partial_{\mu}\Sigma-
{1\over2}\,{\rm tr}\,F_{\mu\nu}F^{\mu\nu}+\cdots,
\end{equation}
where $D_{\mu}=\partial_{\mu}+ig_sG_{\mu}$ is the QCD covariant derivative,
\begin{eqnarray}
V_{\mu}={1\over2}\left(\xi^{\dagger}\partial_{\mu}\xi \,+\,
\xi\partial_{\mu}\xi^{\dagger}\right),\quad
 A_{\mu}={1\over2}i\left(\xi^{\dagger}\partial_{\mu}\xi \,-\,
\xi\partial_{\mu}\xi^{\dagger}\right)
\end{eqnarray}
and $f=F_{\pi}=93~{\rm MeV}$ is the pion decay constant,
\begin{equation}
\Sigma=e^{2i\pi/f}=\xi^2.
\end{equation}
The Nambu-Goldstone fields  $\pi=\pi_a\,T_a$ with
${\rm tr}\left(T_aT_b\right)=1/2\,\delta_{ab}$.
In $\chi$QET the higher-order terms or explicit chiral symmetry breaking
terms are suppressed by powers of $\Lambda_{{\chi}{\rm SB}}\simeq 4\pi\,f$.
The typical size of symmetry breaking terms is
\begin{equation}
{m^2\over \Lambda^2_{\chi{\rm SB}}}\approx 0.1\,.
\end{equation}
Therefore, they
make only small contributions to the naive quark model. Furthermore,
by matching with QCD, one finds the axial coupling and the strong coupling to be
small as well,
$g_A\simeq0.75$ and $\alpha_s\simeq0.28$ so that the perturbation is reliable.
This is why the naive quark model is so successful
in describing non-exotic hadrons~\cite{Manohar:1983md}.

Similarly, the success of diquark picture
for exotic hadrons may be explained by the diquark chiral effective theory
(D$\chi$ET), proposed recently by Sohn, Zahed, and one of
the authors~\cite{Hong:2004xn}.
Since the diquarks have mass around $450~{\rm MeV}$~\cite{Jaffe:2003sg,Hong:2004xn},
they fit naturally in the range where the $\chi$QET is applicable.
Hence, in addition to the relevant degrees of freedom
of $\chi$QET, the diquark chiral effective theory
involves one more degree of freedom, namely a diquark
field, defined as
\begin{equation}
\varphi_{\alpha}^i(x)=\lim_{y\to x}{ \left|y-x\right|^{\gamma_m}\over \kappa^2}
\epsilon^{ijk}\epsilon_{\alpha\beta\gamma}\bar\psi_{cj}^{\beta}(x)\gamma_5
\psi_k^{\gamma}(y),
\end{equation}
where $\kappa$ is a mass scale for the diquark field,
$\gamma_m$ is the anomalous dimension of the diquark correlator
and the Greek indices denote colors, while the Latin indices
$i,j,k=1,2,3$ denote flavors.
$\psi_c\equiv C\bar\psi^T$ is a charge conjugated field of a quark field,
$\psi$, and $C=i\gamma_2\gamma_0$. (Here we consider only scalar diquarks,
since tensor diquarks~\cite{Shuryak:2003zi}
do not contribute to light scalar mesons.)

The diquark chiral effective Lagrangian is described by
\begin{eqnarray}
{\cal L}&=&{1\over2}\left|\left(D_{\mu}+iV_{\mu}\right)
\varphi\right|^2
-{1\over2}M^2\left|\varphi_i\right|^2
-\left(g\,\varphi_i^{\alpha}\,
\epsilon^{ijk}\,\epsilon_{\alpha\beta\gamma}\,
\bar\psi_{cj}^{\beta}\,\gamma_5\psi_k^{\gamma}+{\rm h.c.}
\right)\nonumber\\
& &+\,h_A\,{\rm Tr}\left(\varphi^*\varphi
A_{\mu}^2\right)+h_A^{\prime}\,\varphi^*\varphi\,{\rm Tr}\,A_{\mu}^2\,
+ \,{\cal L}_{\rm int}(\varphi, \psi, \bar\psi)+{\cal L}_{\chi {\rm Q}},
\label{deft}
\end{eqnarray}
where higher-order interactions
of diquarks and quarks is denoted as ${\cal L}_{\rm int}$, which may contain
the (chirally) covariant derivatives. In this paper,
we will take $M=420~{\rm MeV}$ from the random instanton model~\cite{Schafer:1993ra}
and $g^2=3.03$, the best value fitting the scalar
nonet mass~\cite{Hong:2004xn}. The Yukawa coupling splits diquark masses,
$M_{ud}=450~{\rm MeV}$ and $M_{us}=M_{ds}=630~{\rm MeV}$.
And the binding energy between a diquark and an antidiquark in
scalar nonet is $B=280~{\rm MeV}$. We list these parameters
in Table \ref{previous}.

\begin{center}
\begin{table}[htb]
\begin{tabular}{cccccc}
\hline $M_{qs}$(MeV) & $M_{ud}$(MeV) & $B$ (MeV)& $g^2$
\\ \hline
630 & 450 & 280 & 3.03 \\ \hline
\end{tabular}
\caption{The best-fit parameters in \cite{Hong:2004xn}. They
are obtained from the random instanton model and the
scalar nonet masses. }\label{previous}
\end{table}
\end{center}

\section{Diquark-Antidiquark to Goldstone bosons}
\label{int}
\begin{figure}
\epsfxsize=2.5in
\centerline{\epsffile{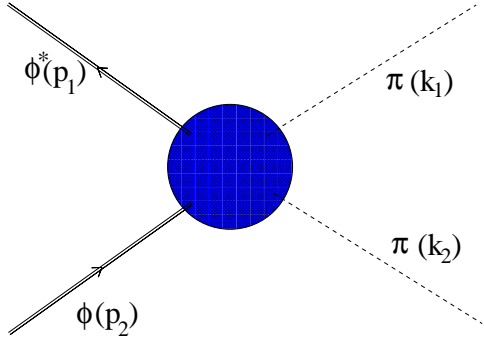}}
\caption{The annihilation process, $\varphi+\varphi^*\to\pi+\pi$.
The double line denotes diquarks and the dashed line denotes pions and kaons.}
 \label{fig1}
\end{figure}

In general, it is very hard to deal with bound states of strongly interacting
particles, especially with broad resonances.
However,  we will attempt to calculate the decay width of strongly interacting
scalar nonet, utilizing  the diquark chiral effective theory.
We will follow the scheme of nonrelativistic positronium decay,
assuming the diquark and NG boson couplings are not large. We will see later
indeed the couplings are small and the naive diquark model
of Jaffe and Wilczek is reliable. In the effective theory, the scalar mesons are
bound states of diquark and anti-diquark.
The decay process occurs as annihilation of diquark with anti-diquark
into NG bosons (see Fig.~\ref{fig1}). The decay width is then given as
\begin{eqnarray}
\Gamma_{S\rightarrow \pi\pi} =
\lim_{v_2\rightarrow 0}v_2\,\sigma\left(\varphi+\varphi^*\to\pi+\pi\right)
|\Psi (0)|^2
\end{eqnarray}
where $v_2$ is the velocity of anti-diquark in the rest frame of
the target diquark and $\Psi(0)$ is the 1S wave function of the
diquark-antidiquark bound state at the origin.
The differential cross section for the annihilation
process is then
\begin{eqnarray}
{\rm d}\sigma = {(2\pi)^4\left|\cal M\right|^2\over
4\sqrt{(p_1\cdot p_2)^2-M_1^2M_{2}^2}}\,{\rm d}\Phi(p_1+p_2;k_1,k_2),
\end{eqnarray}
where $\cal M$ is the annihilation amplitude
for $\varphi^*+\varphi\rightarrow\pi+\pi$.
The phase space is given as
\begin{eqnarray}
{\rm d}\Phi(p_1+p_2;k_1,k_2)=\delta^4(p_1+p_2-k_1-k_2){d^3k_1\over(2\pi)^32E_3}
{d^3k_2\over(2\pi)^32E_4}.
\end{eqnarray}
Integrating over the phase space, we get
\begin{equation}
\Gamma_{S\rightarrow \pi\pi} =
\frac{1}{16\pi}\frac{|{\cal M}|^2}{M_1M_2}
\frac{|\vec{k}_1|}{\sqrt{\vec{k}_1^2+m_A^2}
+\sqrt{\vec{k}_1^2+m_B^2}}|\Psi (0)|^2,
\label{naive}
\end{equation}
where $M_1$ is anti-diquark mass, $M_2$ diquark mass,
$m_A$ the mass of pion $A$, $m_B$ the mass of pion
$B$, and $\vec k_1$ (or $-\vec k_1$)
is the final momentum of pions in the CM frame.

To the order we are interested in
\begin{equation}
{\cal M}={\cal M}_{\rm tree}+{\cal M}_{1\,{\rm loop}}.
\end{equation}

\subsection{Tree contributions}
\begin{figure}
\epsfxsize=4.5in
\centerline{\epsffile{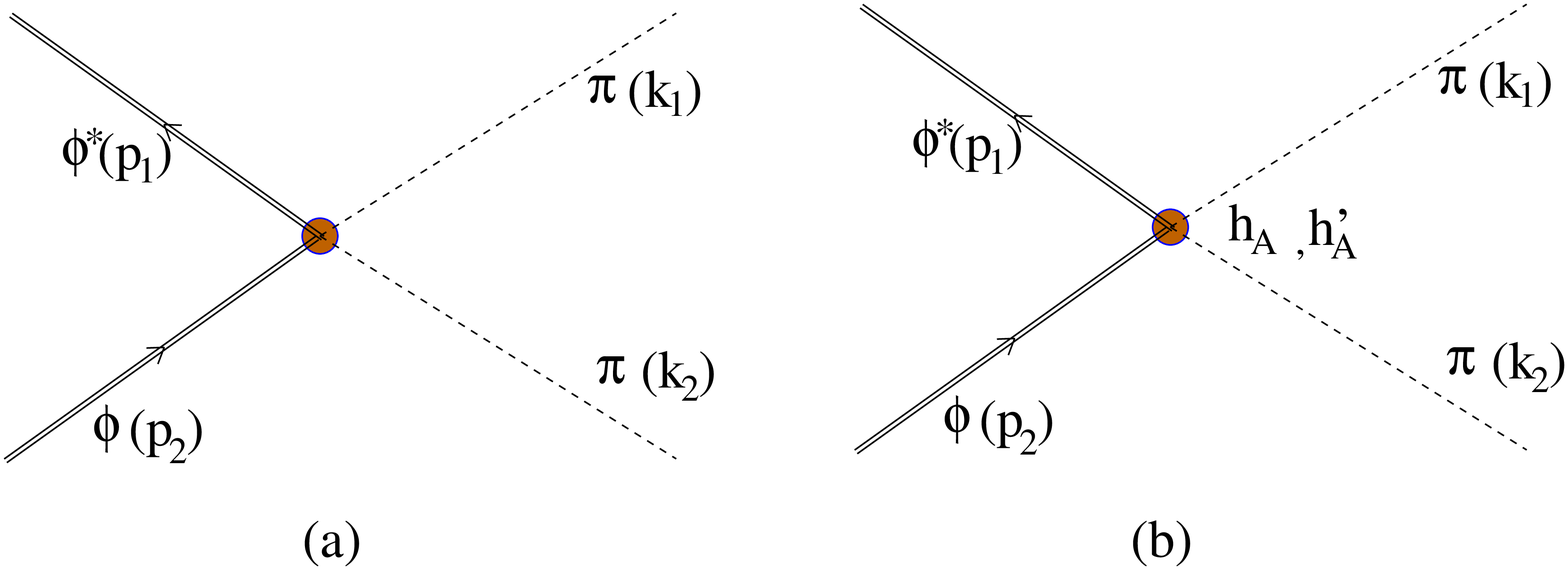}}
\caption{The diquark-pion vertices: (a) vector coupling and (b) axial coupling.   }
 \label{fig2}
\end{figure}

In D$\chi$ET, the leading order contribution to the amplitude
in $1/\Lambda_{\chi\rm SB}$ expansion, shown in Fig.\ref{fig2},
is given as
\begin{eqnarray}
{\cal M}_{\rm tree}=\left<\pi^A(k_1),\pi^B(k_2)\right|\,
{\cal L}_1\,\left|\varphi_a(p_1),\varphi^*_b(p_2)\right>,
\end{eqnarray}
where
\begin{equation}
{\cal L}_1=\,{\rm Tr}\,V^{\mu}\left(\varphi^*\partial_{\mu}\varphi
-\varphi\partial_{\mu}\varphi^*\right)+h_A\,{\rm Tr}\left(\varphi^*\varphi
A_{\mu}^2\right)+h_A^{\prime}\,\varphi^*\varphi\,{\rm Tr}\,A_{\mu}^2\,.
\end{equation}
Note that because of the parity invariance the leading coupling of (scalar)
diquark with pions occurs at $1/\Lambda^2_{{\chi}\rm SB}$.
For the manifest $SU(3)$ covariance, we rewrite the incoming state in the
basis of $SU(3)$ octet and singlet:
\begin{equation}
\left|\varphi_a(p_1),\varphi^*_b(p_2)\right>=2\,T^C_{ab}\,\left|S^C(p_1,p_2)\right>
+{1\over3}\,\delta_{ab}\,\left|S^0(p_1,p_2)\right>.
\end{equation}
Then, the tree amplitudes for octet and singlet  become, respectively,
\begin{eqnarray}
{\cal M}_{\rm tree}^{C\to AB}&=&{i\over4f^2}(k_1-k_2)\cdot(p_1-p_2)\,f^{ABC}-
{h_A\over 2f^2}\,k_1\cdot k_2\,d^{ABC}\\
{\cal M}_{\rm tree}^{0\to AB}&=&-{{h_A+h_A^{\prime}}\over f^2}\,k_1\cdot k_2\,\delta^{AB}.
\end{eqnarray}

\subsection{One-loop contributions}
\begin{figure}[h]
\epsfxsize=2in {\centerline{\epsffile{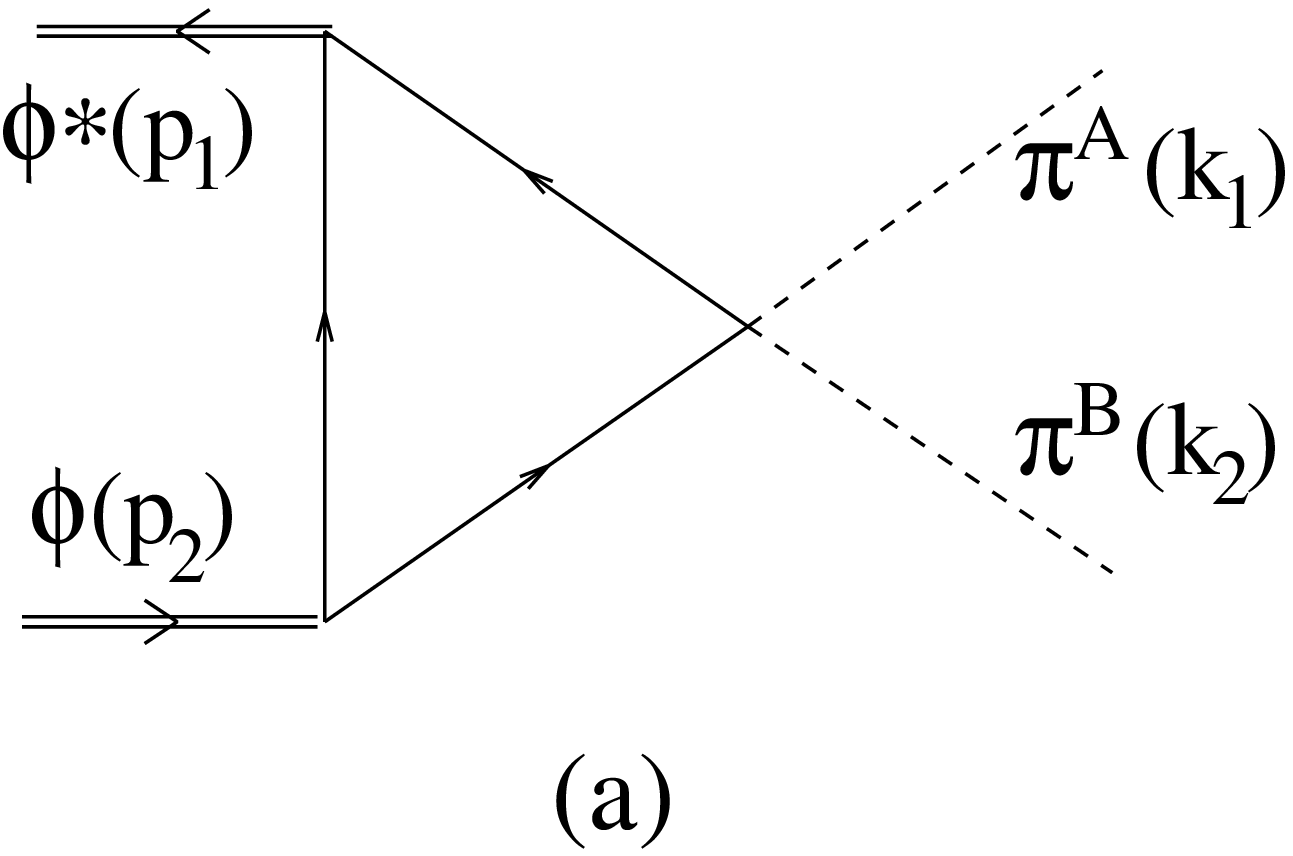}\hskip 1in
{\epsfxsize=2in \epsffile{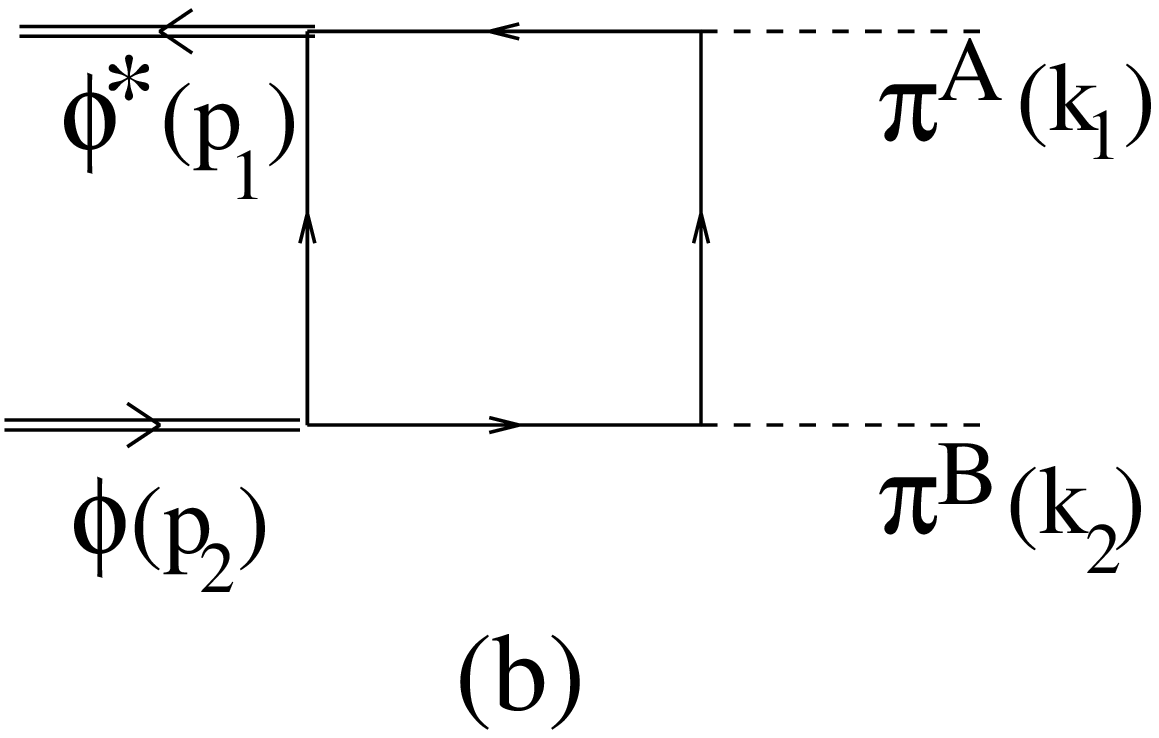}}}} \caption{One loop
amplitudes: (a) vector coupling and (b) axial coupling. Solid
lines denote quark fields.}
 \label{fig3}
\end{figure}

The vector-current of pions (Fig.~\ref{fig3}a) contributes in the
modified minimal subtraction scheme as
\begin{equation}
{\cal M}_{1\,{\rm loop}}=-\frac{6ig^2}{16\pi^2f^2}f^{ABC} T^C_{ba}
\left[ (k_2-k_1)\cdot(p_1-p_2)\,I_1+\cdots\right],
\end{equation}
where the ellipsis denotes the higher-order terms and
\begin{equation}
I_1=\int^1_0dx\int^{1-x}_0(4-3x+3y)
\log\left[\frac{(xp_1-yp_2)^2-xp_1^2-yp_2^2+m^2}{\mu^2}
\right]
\end{equation}
in the SU(3) limit.
The axial-current contribution (Fig.~\ref{fig3}b) becomes
\begin{eqnarray}
{\cal M}_{1\,{\rm loop}}&=&\frac{6g^2g_A^2}{16\pi^2 f^2}\,(k_1\cdot k_2)\,
\big[ \epsilon^{alm}\epsilon^{bon}T^A_{ln}T^B_{mo}
I_2-\{\delta^{AB}\delta_{ab}-2(T^AT^B)_{ba}\}I_3
+\cdots\big]
\end{eqnarray}
where
\begin{eqnarray}
I_2\!\!&=&\!\!6\!\int_0^1\!{\rm d}x\!\int_0^{1-x}\!\!\!{\rm d}y\!\int_0^{1-x-y}
\!\!\!{\rm d}z
\,\log\!\left[\frac{(xp_1+yp_1-yk_1+zk_2)^2-xp_1^2-y(p_1-k_1)^2
-zk_2^2+m^2}{\mu^2}\right]\nonumber\\
I_3\!\!&=&\!\!2\!\int_0^1\!{\rm d}x\!\int_0^{1-x}\!\!\!{\rm d}y\!\int_0^{1-x-y}
\!\!\!{\rm d}z
\,\log\!\left[\frac{(xk_1+yk_1+yk_2+zp_1)^2-xk_1^2-y(k_1+k_2)^2-zp_1^2
+m^2}{\mu^2}\right].
\nonumber\end{eqnarray}

\section{Tetraquark Scalar Decay Widths}
\label{fit}
\subsection{Decays into two nonstrange Nambu-Goldstone bosons}
In our constituent diquark picture, the light scalar mesons have a
binding energy of $280~{\rm MeV}$, which is not negligible in
calculating the decay width. Because of the large binding energy,
Eq.~(\ref{naive}) is far from the proper energy-momentum
conservation.
(The velocity $v_2$ does not also go to zero. However, it factors
out in the leading order.)
To take into account of the binding energy in the
energy-momentum conservation we modify Eq.~(\ref{naive}) as
\begin{equation}
\Gamma_{S\rightarrow \pi^A\pi^B} = \frac{1}{16\pi}\frac{|{\cal
M}|^2}{M_1^*M_2^*} \frac{|\vec{k}_1|}{\sqrt{\vec{k}_1^2+m_A^2}
+\sqrt{\vec{k}_1^2+m_B^2}}|\Psi (0)|^2\,, \label{naive2}
\end{equation}
where
$M_1^*=M_1-BM_1/(M_1+M_2)$, $M_2^*=M_2-BM_2/(M_1+M_2)$, and
\begin{equation}
\vec{k}_1^2=\frac{\{
(M_1^*+M_2^*)^2+m_B^2-m_A^2\}^2}{4(M_1^*+M_2^*)^2} -m_B^2.
\end{equation}

By a naive dimensional analysis we find
\begin{equation}
|\Psi (0)|^2=c(MB)^{3/2}
\end{equation}
where $c$ is a constant and
$M$ is the reduced diquark mass. From the analogy with
the 1S wavefunction of positronium, we take $c=1/\pi$ here.

The lightest scalar $\sigma$ is believed to have no or just little
strangeness in it.
So we assume the ideal mixing between $\sigma$ and $f_0$.
In our calculation, the
(anti)diquark in $\sigma$ is nonstrange and the (anti)diquark in
$f_0$ includes one strange (anti)quark. We take $g_A=0.75$, the
experimental value in the quark
model~\cite{Manohar:1983md,Weinberg:1991gf}, and set the
quark mass $m= 400~{\rm  MeV}$~\footnote{ We assume SU(3) symmetric
mass. The SU(3) breaking mass terms contribute less than 10 percent.}.
We then determine the couplings $h_A$
and $h_A^{\prime}$ by fitting the experimental values for the decay
widths of scalar mesons in the dominant modes. The results are
shown in Table \ref{width1}~\footnote{Since the pions in the final
state carry large momenta, a fraction of $\Lambda_{{\chi}SB}$,
one has to include the re-scattering effect of pions in
the final state. However, we find that such effect is about a few
percents and thus we neglect it in our analysis.}. We find our
couplings are of order 1 for the decay widths in the range allowed
by experiments. Since the expansion parameter is $h_A^2/4\pi$ or
$h_A^{\prime2}/4\pi$ rather than the coupling itself, the
higher-order corrections are quite suppressed and the perturbation
is reliable.
\begin{center}
\begin{table}[htb]
\begin{tabular}{cccccc}
\hline  set & $h_A$ & $h_A^{\prime}$ &
~$\Gamma_{\sigma\rightarrow 2\pi}$(MeV)~&
~$\Gamma_{f_0\rightarrow 2\pi}$(MeV)~ &
~$\Gamma_{a_0\rightarrow \eta\pi}$(MeV)~
\\ \hline
1 & ~1.46~ &~ $-1.66$~ & ~317~ & ~67~ & ~83\\ \hline
2 & ~1.20~ &~ $-1.38$~ & ~234~ & ~71~ & ~57 \\ \hline
\end{tabular}
\caption{Two fitting parameters and the decay widths of
the light scalar mesons in dominant modes. }\label{width1}
\end{table}
\end{center}
Once we determine the axial couplings $h_A$ and $h_A^{\prime}$, we
calculate in the next section the decay width of $f_0$ in its subdominant
mode to see if D$\chi$ET works.

\subsection{$f_0$ decay into two kaons}
Since the mass of two kaons is larger than $f_0$ mass,
we cannot obtain the decay width with Eq.~(\ref{naive2}).
The broadening of $f_0$ makes the decay into two kaons
possible. So the broadening should be considered in order to
calculate the decay mode into two kaons.
Keeping the basic picture, we introduce the Breit-Wigner form
to revise Eq.~(\ref{naive2});
\begin{equation}
\Gamma_{S\rightarrow \pi^A\pi^B} =
N\int_{s_{min}}^{s_{max}}{\rm d}s\,\frac{|{\cal M}|^2}{4\pi^2\,s}
\frac{|\vec{k}_f|}{\sqrt{\vec{k}_f^2+m_A^2}
+\sqrt{\vec{k}_f^2+m_B^2}}\,
\frac{m_S\,\Gamma_{S}\,|\Psi_s (0)|^2}{(s-m_S^2)^2
+m_S^2\Gamma_{S}^2}
\label{res}
\end{equation}
with scalar mass $m_S$ and total decay width $\Gamma_S$.
$\sqrt{s_{max}}$ is $m_S+\Gamma_{S}$ and
$\sqrt{s_{min}}$ is chosen to be the larger one between the threshold
value for the decay mode and $m_S-\Gamma_{S}$.
So $\sqrt{s_{min}} = m_K +m_{\bar{K}}$ for $f_0\rightarrow
K+\bar{K}$ mode. The wavefunction at the origin and the final
momentum are modified as
\begin{equation}
|\Psi_s (0)|^2\approx \{ (M_1+M_2-\sqrt{s}) M\}^{3/2}/\pi
\end{equation}
and
\begin{equation}
\vec{k}_f^2=\frac{( s+m_B^2-m_A^2)^2}{4s}
-m_B^2
\end{equation}
in (\ref{res}).
The normalization constant $N$ is adjusted by
\begin{equation}
\frac1N=\int_{s_{min}}^{s_{max}}ds\frac{1}{\pi}
\frac{m_s\Gamma_{S}}{(s-m_S^2)^2
+m_s^2\Gamma_{S}^2}.
\end{equation}
Since $f_0\rightarrow 2\pi$ is dominant among $f_0$ decay modes,
we will use $\Gamma_{f_0\rightarrow 2\pi}$ in Table \ref{width1} for
$\Gamma_{S}$ in Eq. (\ref{res}). Then we obtain
75 MeV with set 1 and 59 MeV with set 2 for
$\Gamma_{f_0\rightarrow 2K}$.
Experimentally
$\Gamma_{f_0\rightarrow 2K}/(\Gamma_{f_0\rightarrow 2\pi}+
\Gamma_{f_0\rightarrow 2K})$ is 0.14 $\sim$ 0.32 and the full decay
width of $f_0$ is $\Gamma=40\sim100~{\rm MeV}$~\cite{Hagiwara:2002fs}.
We find our result is a few times larger than the experimental value,
$\Gamma_{f_0\rightarrow 2K}=5.6\sim 32 ~{\rm MeV}$.
Considering the experimental uncertainty and the crude approximation
we made for the bound state wavefunction, our result is not too bad.

\section{Conclusion}
We calculate the decay width of light scalar mesons in the framework
of the diquark chiral effective theory, which was introduced recently
to describe exotic hadrons like pentaquarks. Using the parameters fixed
previously by the mass of scalar mesons, we determine the axial couplings
of diquarks with pions and kaons from the experimental data for the
decay width of scalars in the dominant decay modes. The axial couplings
turned out to be small, showing that the naive diquark model of
Jaffe and Wilczek is reliable. Then, we calculate the decay width of
scalar mesons for the subdominant channel and find that our result is
not far off from the experimental data.

To conclude, in the diquark chiral effective theory, based on the
spontaneously broken chiral symmetry and the diquark picture of exotic
hadrons, we find the decay width of scalar mesons in good agreement of
experimental data. It shows that the diquark chiral effective theory
captures correct physics of exotic hadrons.

\newpage


\acknowledgments
We thank P. Ko, F. Sannino and Y. J. Sohn for useful
comments. The work of D.K.H. is supported by Korea Research
Foundation Grant (KRF-2003-041-C00073). The work of C.S. was
financially supported by Pusan National University in the program,
Post-Doc.~2004.

\end{document}